\documentclass[twocolumn,pra,showpacs,aps]{revtex4-1}
\usepackage{amssymb}
\usepackage{amsmath}
\usepackage{graphicx}
\usepackage{epsfig}

\begin{document}

\title{Helical resonant transport and purified amplification at an exceptional
point}
\author{K. L. Zhang}
\author{L. Jin}
\email{jinliang@nankai.edu.cn}
\author{Z. Song}
\email{songtc@nankai.edu.cn}
\affiliation{School of Physics, Nankai University, Tianjin 300071, China}

\begin{abstract}
We propose an application of a parity-time symmetric non-Hermitian
Su-Schrieffer-Heeger (SSH) model by embedded it in a two-dimensional square
lattice tube. The coalescence state at the exceptional point of
non-Hermitian SSH model is chiral and selectively controls helical transport
and amplification. Two typical helicity-dependent scattering dynamics are
observed. If the incidence has an identical helicity with the embedded
non-Hermitian SSH model, we observe a perfect transmission without
reflection. However, if the incidence has an opposite helicity with the
embedded non-Hermitian SSH model, except for a full transmission, we observe
an amplified transmission with different helicity from the incidence; but
the amplified reflection has identical helicity with the incidence. These
intriguing features are completely unexpected in Hermitian system. Moreover,
the helical amplification at high efficiency can be triggered by an
arbitrary excitation. The different dynamics between incidences with
opposite helicities are results of unidirectional tunneling, which is
revealed to be capable of realizing without introducing magnetic field. 
We give a discussion about the helical dynamics under system imperfections.
Our findings open a direction in all-optical device and provide perspectives in
non-Hermitian transport.
\end{abstract}

\maketitle

\section{Introduction}

Nowadays non-Hermitian physics has emerged as a versatile platform for the
exploring of functional devices that are absent or difficult to realize in
Hermitian regime \cite{PRL08a,PRL08b,Klaiman,CERuter,YDChong,Regensburger,LFeng,Longhi15,Fleury,BenderRPP,
	NM,FL,Ganainy18,YFChen,Christodoulides,LonghiEPL, MANiri, Ozdemir}. The
passive and active $\mathcal{PT}$-symmetric non-Hermitian systems with
balanced gain and loss are investigated theoretically and experimentally in
various setups in optics \cite%
{Ruschhaupt,OL07,AGuo,HJingPRL2014,BPeng,LChang,LFengScience,HodaeiScience}.
The non-Hermiticity induces nonunitary dynamic including the power
oscillation \cite{Klaiman,CERuter} and the unidirectional invisibility \cite%
{Regensburger,LFeng,Longhi15}. However, the $\mathcal{PT}$ symmetry protects
the symmetry of transmission \cite{Muga,AnnPhys,Ahmed,Ali4}. An ideal
building block possessing asymmetric transmission is the asymmetric dimer
with an unequal hopping strength \cite{ZXZAnn}.
Asymmetric hopping strength can be realized in coupled microring resonators
with synthetic imaginary gauge fields in photonics \cite{LonghiSR, LonghiPRB}.
Moreover, asymmetric transmission is possible under gain and loss associated
with effective magnetic flux \cite{LXQ,Longhi,JLSR,JLSR2,CLi17,JLNJP,JLPRL}.
Nonreciprocal photonics in the non-Hermitian physics are revealed to be
useful for the applications in optics \cite%
{JLPRL,AAClerk,Koutserimpas,HRamezani,Huang,LonghiOL}.

Non-Hermitian system at exceptional point (EP) has a coalescence state \cite%
{Midya,MANiri}. We propose an application of the $\mathcal{PT}$%
-symmetric non-Hermitian Su-Schrieffer-Heeger (SSH) model at the EP \cite%
{Poli,Weimann,Pan,Jean,Zhao,Parto,BXWang}, which exhibits unidirectionality without
the assistance of magnetic field. The unidirectionality leads to the
helicity dependent dynamics of wave propagation. The non-Hermitian SSH ring
is embedded in a square lattice tube center as a scattering center. The wave
propagation of different incidences are studied. The non-Hermitian SSH ring
at EPs has a coalescence state, which is chiral due to the chirality of EP
and coincides with one of the two degenerate zero modes of a uniform ring.
Thus, the non-Hermitian SSH ring engineered at the EPs results in helical
transport. The non-Hermitian SSH ring at the EP leads to two typical
helicity-dependent scattering behaviors: (i) zero transverse bandwidth
perfect transmission without reflection; (ii) amplified zero transverse
bandwidth interfered transmission with reflection. In the later case, except
for a resonant transmission, part of the transmission after scattering has
opposite helicity with the incidence; and the reflection has identical
helicity with the incidence. This dramatically differs from the traditional
helical scattering that transmission after scattering has identical helicity
with the incidence, but the reflection has opposite helicity with the
incidence. These features are useful for the control of helical transport
and amplification. We also show that the helical amplification can be
triggered by an arbitrary excitation at high efficiency.

This paper is organized as follows. In Sec. \ref{model and helical
scattering center}, we present the square lattice tube model embedded with a
non-Hermitian SSH ring. In Sec. \ref{Unidirectional hopping}, we discuss the
unidirectional hopping and show the helical dependent dynamics under the
influence of the non-Hermitian SSH ring. In Sec. \ref{dynamics}, we
demonstrate the purification and amplification for a single-site excitation.
In Sec. \ref{Discussion}, we investigate the system with imperfections and
discuss the possible experimental realization. In Sec. \ref{Summary}, we
summarize our findings.

\section{Chiral scattering center}

\label{model and helical scattering center}

\begin{figure}[tbp]
\centering\includegraphics[width=0.48\textwidth]{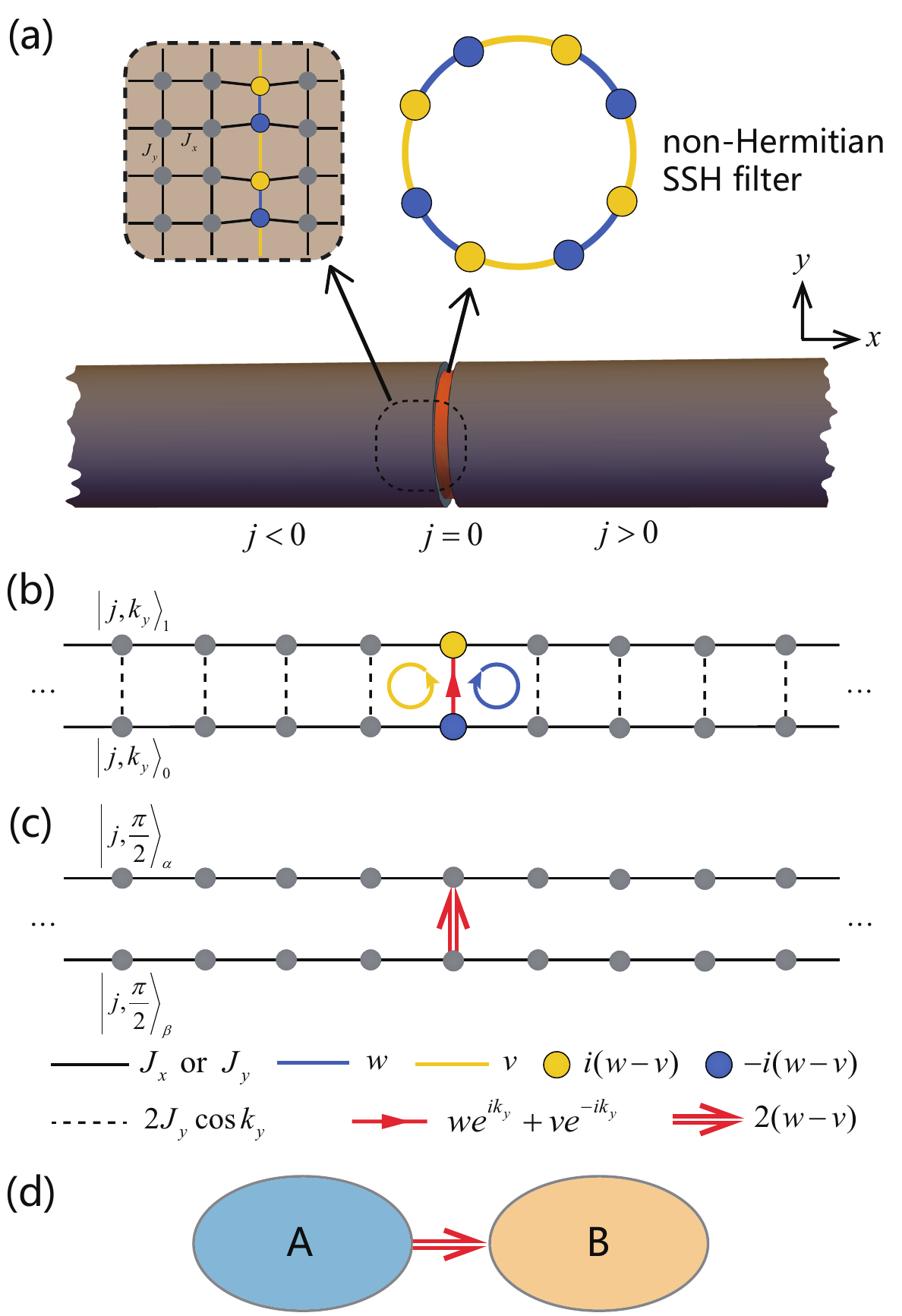}
\caption{(a) Schematic of the infinite lattice tube with a non-Hermitian SSH
ring embedded in the center. The SSH ring is set at the EP. (b) Ladder
lattice of the Hamiltonian [Eq.~(\protect\ref{ladder})] in the $k_y$
momentum space. The scattering center generates two opposite magnetic fluxes
in the two neighbor plaquettes. (c) Schematic of the equivalent ladder of
Eq.~(\protect\ref{H_pi2}). (d) Schematic of two Hermitian subsystems \textrm{A} and \textrm{B} with unidirectional hopping (red double arrow). (c) is a
concrete example of (d).} \label{figure1}
\end{figure}

We consider a non-Hermitian engineering of the tube, the
schematic is illustrated in Fig.~\ref{figure1}(a). A non-Hermitian SSH ring
is embedded in the center ($j=0$) of the tube \cite{Pan,Jean,Parto}. The
non-Hermitian SSH ring has staggered coupling along the $y$ direction, and
the gain and loss are staggered. The dynamics in the tube can be considered
as a scattering problem. Before we handling the problem of chiral scattering
center, we first discuss a uniform square lattice tube without any impurity;
and then discuss the non-Hermitian SSH model and its zero energy eigenstates
at the EP. 

The uniform infinite square lattice tube without any impurity
has infinite sites in the $x$ direction, and the supported momentum in the $%
x $ direction is continuous in the region $k_{x}\in \left[ -\pi ,\pi \right]
$. The tube is periodic along the $y$ direction and has total $2M$ sites,
and the supported momentum in the $y$ direction is discrete $k_{y}=n\pi /M$
(integer $n=1,2,...,2M$). The (unnormalized) eigenstate of the
uniform square lattice tube without any impurity is given by%
\begin{equation}
\left\vert \Psi \right\rangle =\frac{1}{\sqrt{2M}}\sum_{j=-\infty }^{\infty
}e^{ik_{x}j}\sum_{l=1}^{2M}e^{ik_{y}l}\left\vert j,l\right\rangle ,
\end{equation}%
which is a plane wave steady-state scattering solution in the $x$
direction. In the eigenstate of the uniform square lattice tube $\left\vert
\Psi \right\rangle $, $\left\vert j,l\right\rangle $ is the Wannier
representation \cite{WS09,WS13}; $j$ ($l$) is the index of the lattice in
the $x$ ($y$) direction; and $\left\vert j,l\right\rangle $ stands for the
Wannier state localized at lattice site $\left( j,l\right) $ of the lattice
tube.

\begin{figure}[tbp]
\centering\includegraphics[width=0.48\textwidth]{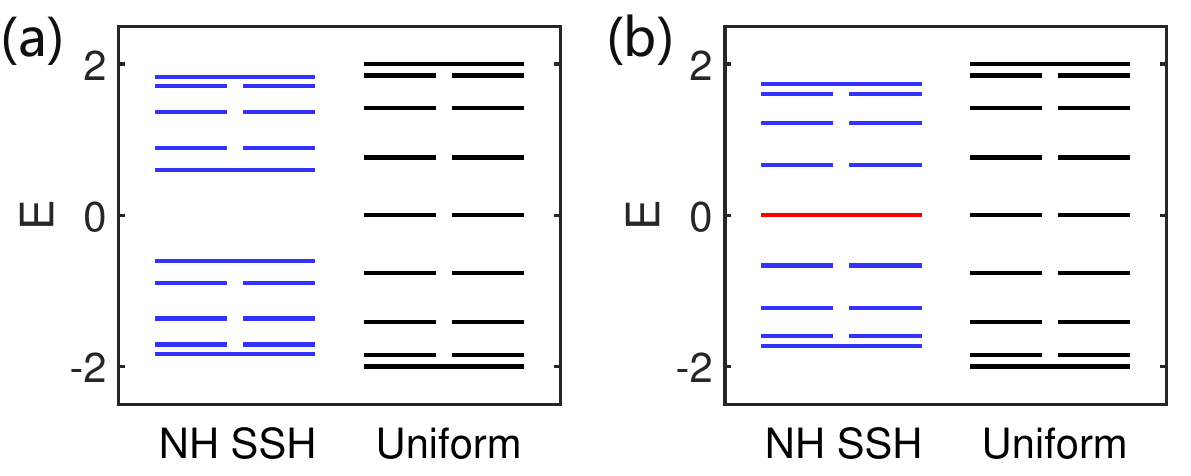}
\caption{Energy levels (solid lines) of the non-Hermitian SSH ring and the
uniform ring. The non-Hermitian SSH ring has staggered coupling $w$ and $v$;
and stagger gain and loss $\pm i\protect\gamma$. The uniform ring has
uniform coupling $J$. (a) $\protect\gamma=0.8$. (b) $\protect\gamma=1$.
Other parameters are $w=1.5$, $v=0.5$, and $J=1$. Both ring sizes are $16$.
The red line in (b) represents the coalesced energy level, other levels in
(b) and all levels in (a) are all mismatched.} \label%
{figure2}
\end{figure}

The eigen energies of the non-Hermitian SSH ring are significantly different
from the eigen energies of the uniform ring. The energy levels of the
non-Hermitian SSH ring and the uniform ring are compared in Fig. \ref%
{figure2}(a); correspondingly, the supported discrete momenta $k_{y}$ are
mismatched. Consequently, the incidences in the tube can not pass the
non-Hermitian SSH ring center and are blocked. However, if the non-Hermitian
SSH ring $H_{\mathrm{SSH}}$ is engineered at the EP, it has a coalescence
eigenstate with zero energy. The coalesced zero mode is%
\begin{equation}
\left\vert \psi _{-}\right\rangle =\frac{1}{\sqrt{2M}}\sum_{l=1}^{2M}e^{-il%
\pi /2}\left\vert 0,l\right\rangle .
\end{equation}%
The intriguing feature of the coalescence eigenstate $\left\vert \psi
_{-}\right\rangle $ is that it is identical to one of the two degenerate
zero energy eigenstates of the uniform ring (at the momentum $k_{y}=\pm \pi
/2$) as elaborated in Fig. \ref{figure2}(b). For the Hermitian conjugation
SSH ring $H_{\mathrm{SSH}}^{\dagger }$, it differs from $H_{\mathrm{SSH}}$
only in the sense that the gain and loss are interchanged. $H_{\mathrm{SSH}%
}^{\dagger }$ is also at the EP and supports a coalescence zero energy
eigenstate in the form of
\begin{equation}
\left\vert \psi _{+}\right\rangle =\frac{1}{\sqrt{2M}}\sum_{l=1}^{2M}e^{+il%
\pi /2}\left\vert 0,l\right\rangle .
\end{equation}%
Notably, the two coalescence eigenstates in $H_{\mathrm{SSH}}$ and $H_{%
\mathrm{SSH}}^{\dagger }$ are the two degenerate zero modes of the uniform
ring, being orthogonal $\langle \psi _{-}\left\vert \psi _{+}\right\rangle
=0 $. The two coalescence eigenstates possess opposite chiralities
classified by the type of EP \cite{Midya,MANiri,Heiss,CTChan,LJinEP}. The
non-Hermitian SSH ring $H_{\mathrm{SSH}}$ engineered at the EP only supports
the resonant transmission for the incident wave with one of two momenta $%
k_{y}=\pm \pi /2$.

The non-Hermitian SSH ring in the tube is regarded as the scattering center;
the semi-infinite tubes coupled to the non-Hermitian SSH ring are regarded
as the leads. Notably, the leads are translational invariant by shifting
every one site in the $y$ direction; however, the non-Hermitian SSH ring
center is translational invariant by shifting every two sites in the $y$
direction. This is an important feature for the design of helical transport
in this paper.

To comprehensively address the scattering inside the tube, we write down the
tube Hamiltonian in the momentum space $H=\sum_{k_{y}}H_{k_{y}}$. $H_{k_{y}}$%
\ describes a ladder system as schematically illustrated in Fig. \ref%
{figure1}(b). The Hamiltonian in the momentum subspace $k_{y}$ reads%
\begin{eqnarray}
H_{k_{y}} &=&J_{x}\sum_{j=-\infty }^{\infty }\sum_{\lambda =1,0}\left\vert
j-1,k_{y}\right\rangle _{\lambda \lambda }\left\langle j,k_{y}\right\vert
\notag \\
&&+2J_{y}\cos k_{y}\sum_{j\neq 0,j=-\infty }^{\infty }\left\vert
j,k_{y}\right\rangle _{10}\left\langle j,k_{y}\right\vert  \notag \\
&&+\left( we^{ik_{y}}+ve^{-ik_{y}}\right) \left\vert 0,k_{y}\right\rangle
_{10}\left\langle 0,k_{y}\right\vert +\mathrm{H.c.}  \notag \\
&&-i(w-v)(-1)^{\lambda }\sum_{\lambda =1,0}\left\vert 0,k_{y}\right\rangle
_{\lambda \lambda }\left\langle 0,k_{y}\right\vert ,  \label{ladder}
\end{eqnarray}%
The Fourier transformations applied to the square lattice tube is $%
\left\vert j,k_{y}\right\rangle _{\lambda
}=M^{-1/2}\sum_{l=1}^{M}e^{ik_{y}(2l-\lambda )}\left\vert j,2l-\lambda
\right\rangle $, where $\lambda =1,0$ and $k_{y}=n\pi /M$, $(n=1,2,...,M)$. $%
H_{k_{y}}$ with different momenta commute%
\begin{equation}
\lbrack H_{k_{y}},H_{k_{y}^{\prime }}]=0.  \label{comm relation}
\end{equation}%
This means that $H$\ can be decomposed into $M$-fold independent
sub-Hamiltonians. The ladder $H_{k_{y}}$ has a non-Hermitian dimer embedded
in the center. Remarkably, any distortion $w\neq v$ generates two opposite
effective magnetic fluxes in the two plaquettes with gain and loss.

An interesting situation is when the ladder has zero effective magnetic
fluxes at $k_{y}=\pi /2$ although the presence of nonreciprocal Hermitian
coupling $i\left( w-v\right) $ with a Peierls phase factor $i=e^{i\pi /2}$.
This structure is equivalent to the ladder with asymmetric coupling in Fig. %
\ref{figure1}(c) without employing effective magnetic flux \cite%
{JLPRB,ZKLPRB}. The asymmetric coupling is physical, the
asymmetric coupling amplitudes for photons tunneling in the opposite
directions is proposed in the coupled ring resonator array \cite%
{LonghiSR,LonghiPRB}. From the analysis of the $k_{y}=\pi /2$ subspace, we
can obtain the previous conclusion discussed.

\section{Unidirectional tunneling and helical resonant transport}

\label{Unidirectional hopping}

In the $k_{y}=\pi /2$ subspace, term with coupling $2J_{y}\cos k_{y}$
vanish, and the ladder Hamiltonian in Eq. (\ref{ladder}) reduces to the form
of%
\begin{eqnarray}
H_{\pi /2} &=&J_{x}\sum_{j=-\infty }^{\infty }\sum_{\lambda =1,0}\left\vert
j-1,\pi /2\right\rangle _{\lambda \lambda }\left\langle j,\pi /2\right\vert
\notag \\
&&+i\left( w-v\right) \left\vert 0,\pi /2\right\rangle _{10}\left\langle
0,\pi /2\right\vert +\mathrm{H.c.}  \notag \\
&&-i(w-v)(-1)^{\lambda }\sum_{\lambda =1,0}\left\vert 0,\pi /2\right\rangle
_{\lambda \lambda }\left\langle 0,\pi /2\right\vert .  \label{H_pi1}
\end{eqnarray}%
Taking the unitary transformation
\begin{eqnarray}
\left\vert j,\pi /2\right\rangle _{\alpha } &=&\frac{1}{\sqrt{2}}\left(
\left\vert j,\pi /2\right\rangle _{1}-\left\vert j,\pi /2\right\rangle
_{0}\right) ,  \notag \\
\left\vert j,\pi /2\right\rangle _{\beta } &=&\frac{-i}{\sqrt{2}}\left(
\left\vert j,\pi /2\right\rangle _{1}+\left\vert j,\pi /2\right\rangle
_{0}\right) ,  \label{tran_1}
\end{eqnarray}%
with $j=0,\pm 1,\pm 2,...,$ the Hamiltonian can be rewritten in the form of
\begin{eqnarray}
H_{\pi /2} &=&J_{x}\sum_{j=-\infty }^{\infty }(\left\vert j-1,\pi
/2\right\rangle _{\alpha \alpha }\left\langle j,\pi /2\right\vert  \notag \\
&&+\left\vert j-1,\pi /2\right\rangle _{\beta \beta }\left\langle j,\pi
/2\right\vert )+\mathrm{H.c.}  \notag \\
&&+2\left( w-v\right) \left\vert 0,\pi /2\right\rangle _{\alpha \beta
}\left\langle 0,\pi /2\right\vert ,  \label{H_pi2}
\end{eqnarray}%
which is schematically illustrated in Fig. \ref{figure1}(c). We note that
the non-Hermiticity of $H_{\pi /2}$\ only arises from the unidirectional
hopping term $2\left( w-v\right) \left\vert 0\right\rangle _{\alpha
}\left\langle 0\right\vert _{\beta }$ and $H_{\pi /2}$\ still has parity
symmetry. The composed asymmetric coupling takes the advantages of the two
leads structure, which differs from other cases in the literatures \cite%
{LXQ,Longhi,JLSR,JLSR2,CLi17,JLNJP,JLPRL}.

\begin{figure}[tbp]
\centering
\includegraphics[width=0.4\textwidth]{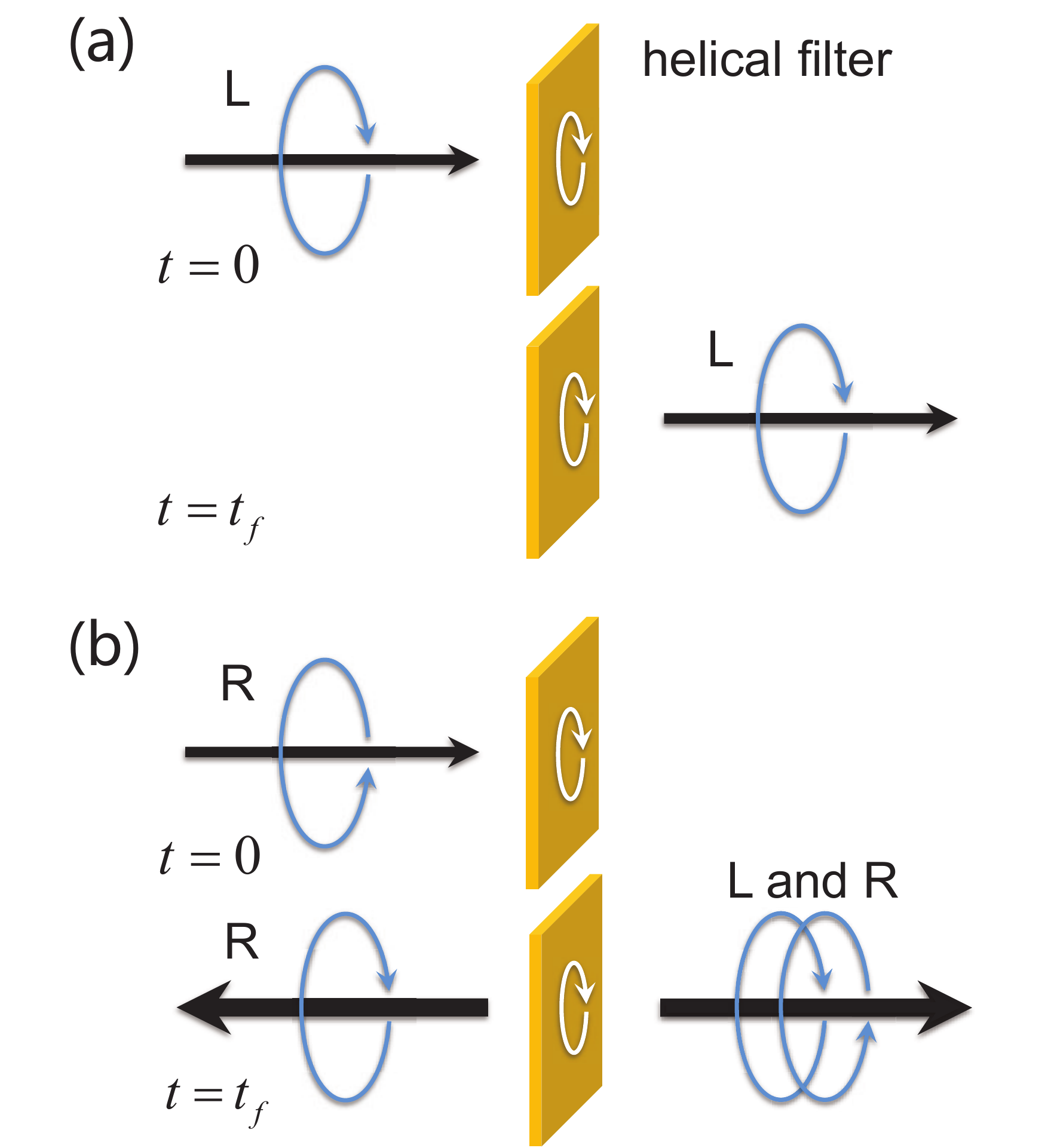}
\caption{Schematic of the helical transport. The helical filter is the SSH
ring at EP. (a) Perfect transmission with zero transverse bandwidth for
incident wave possessing identical helicity with the SSH ring helical
filter. (b) Full transmission associating with out going wave for incidence
possessing opposite helicity with the SSH ring helical filter.} \label%
{figure3}
\end{figure}

\begin{figure*}[tbh]
\centering
\includegraphics[width=1\textwidth]{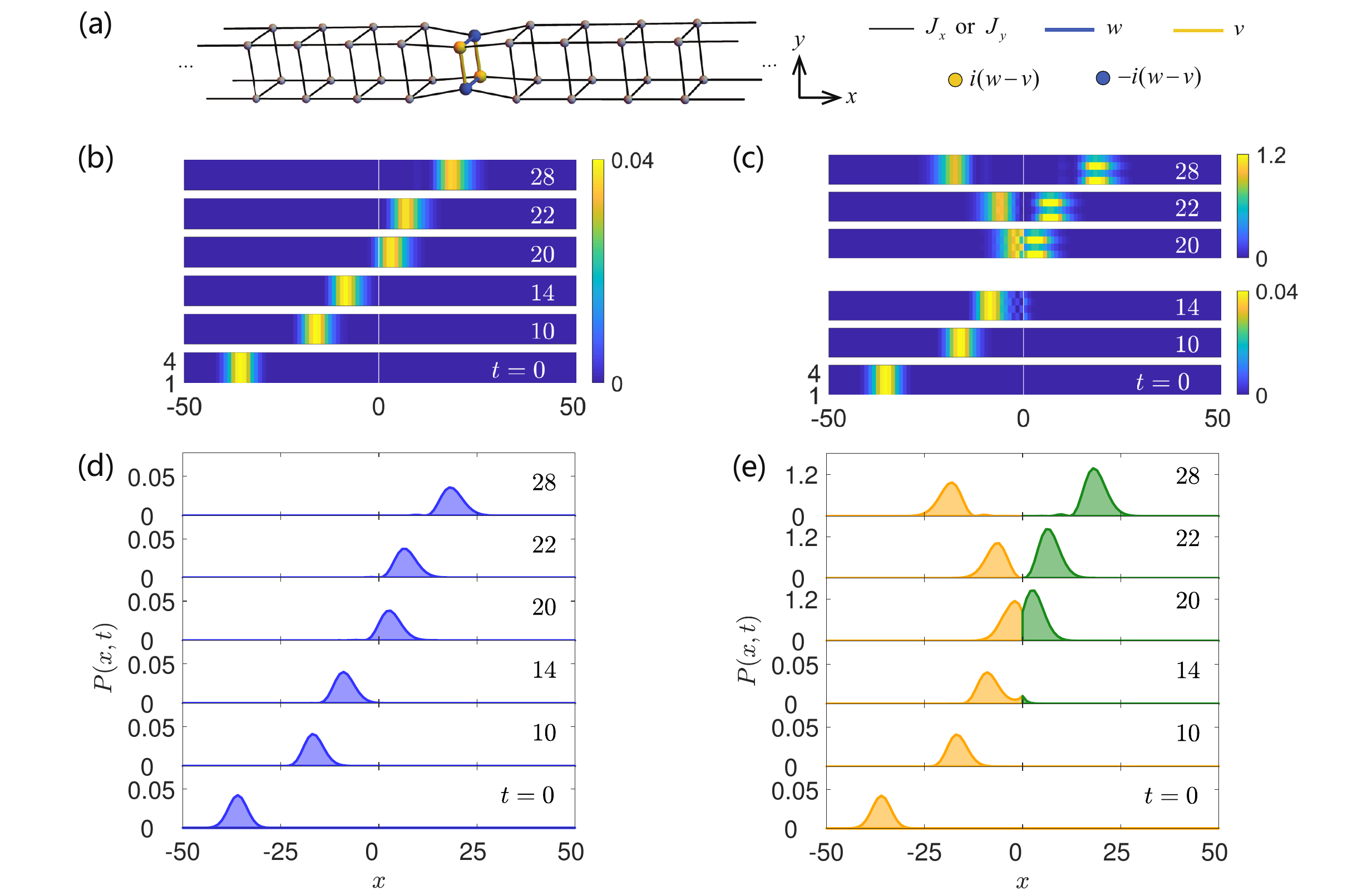}
\caption{(a) Non-Hermitian lattice tube with $M=2$. Snapshots of the
intensity at various time moments for two typical initial excitations: (b)
Perfect transmission of Gaussian profile $\left\vert \Psi _{\mathrm{G}
}^{-}(0)\right\rangle$ with identical helicity of the filter; (c) Gaussian
profile $\left\vert \Psi _{\mathrm{G}}^{+}(0)\right\rangle$ with opposite
helicity of the filter. The interference pattern at $x>0$ in (c) shows light
and dark areas, which indicate the interference of transmitted excitation
with opposite helicities [see Eq. (\protect\ref{pattern})]. (d) and (e) are
the intensity distribution in the $y$ direction for (b) and (c),
respectively. The blue (gold) area represents the output with the identical
(opposite) helicity of the filter, while the green area represents the
output with two helicities. Other parameters of the initial excitation are $%
\protect\alpha _{\mathrm{w}}=0.3$, $k_{x}=-\protect\pi/2$ and $N_{\mathrm{c}
}=-35$ and for the system are $J_{x}=J_{y}=J=0.25$, $w=1.5$, $v=0.5$ and the
time is in unit of $J^{-1}$. }
\label{figure4}
\end{figure*}

A tight-binding network is constructed topologically by the sites and
various connections between them. There are three types of basic
non-Hermitian clusters leading to the non-Hermiticity of a discrete
non-Hermitian\ system: i) complex on-site potential denoted as $\left(
V+i\gamma \right) \left\vert l\right\rangle \left\langle l\right\vert $; ii)
non-Hermitian dimer denoted as $e^{i\varphi }(\left\vert l\right\rangle
\left\langle j\right\vert +\left\vert j\right\rangle \left\langle
l\right\vert )$, and iii) asymmetric hopping amplitude dimer denoted as $\mu
\left\vert l\right\rangle \left\langle j\right\vert +\nu \left\vert
j\right\rangle \left\langle l\right\vert $ ($\mu \neq \nu $), where $%
V,\gamma ,\varphi ,\mu ,\nu $\ are real numbers. The asymmetric hopping
induces imaginary magnetic flux and has been used in modeling a
delocalization phenomenon \cite{Hatano}. The unidirectional hopping is
defined as $\mu \nu =0$, which is an extreme non-Hermitian term, only allows
the particle tunneling from \textrm{A} to \textrm{B}.

The unidirectional hopping we encounter here is a basic non-Hermitian
element in the context of tight-binding network, which leads to
unidirectional tunneling between two Hermitian subsystems \textrm{A} and
\textrm{B} [Fig. \ref{figure1}(d)]: On the one hand, when an initial state
is set in subsystem \textrm{B}, the particle is always confined in \textrm{B}
and thus the Dirac probability is conservative. On the other hand, when an
initial state is set in subsystem \textrm{A}, the particle can tunnel to
\textrm{B} and thus the Dirac probability is not conservative. Subsystem can
be regarded as a conditional invariant subspace, which is an exclusive
feature of a non-Hermitian system. These features can be seen by considering
the scattering problem of Hamiltonian Eq. (\ref{H_pi2}).

We take the momentum $k_{y}=\pm \pi /2$ as an inner degree of freedom; $%
k_{y}=-\pi /2$ indicates the circling toward the positive direction of $y$,
thus, an angular momentum associated with $k_{y}=-\pi /2$ is toward the
negative direction of $x$; in contrast, an angular momentum associated with $%
k_{y}=\pi /2$ is toward the positive direction of $x$. For incidence with $%
k_{y}=-\pi /2$ ($k_{y}=\pi /2$) moving toward the positive direction of $x$,
the direction of angular momentum associated with $k_{y}$ is opposite
(identical) to its propagation direction along $x$, thus, we refer to the
helicity as the left-handed (right-handed).

Now we consider the scattering problem for a unidirectional\ scattering
center, which is exactly solvable and is of significant not only for the
non-Hermitian physics but also for applications in optics. A direct
motivation is that the equivalent Hamiltonian of the present Hamiltonian in
an invariant subspace is a concrete example with unidirectional tunneling.
Two solutions of the scattering wave function in the subspace $H_{\pi /2}$
can be obtained by the Bethe ansatz method (see Appendix A). For incidence
with momentum $k_{x}$ in the $x$ direction, the resonant transmission
solution is\
\begin{equation}
\left\vert \psi _{k_{x}}^{L}\right\rangle =\frac{1}{\sqrt{2M}}%
\sum_{j=-\infty }^{\infty }\sum_{l=1}^{M}e^{ik_{x}j}(-1)^{l}\left(
\left\vert j,2l-1\right\rangle -i\left\vert j,2l\right\rangle \right) ,
\label{solution1}
\end{equation}%
and the interfered transmission solution is%
\begin{eqnarray}
\left\vert \psi _{k_{x}}^{R}\right\rangle &=&\frac{1}{\sqrt{2M}}%
\sum_{j=-\infty }^{\infty }\sum_{l=1}^{M}[e^{ik_{x}j}(-1)^{l}\left(
\left\vert j,2l-1\right\rangle +i\left\vert j,2l\right\rangle \right)  \notag
\\
&&+A_{k_{x}}e^{ik_{x}\left\vert j\right\vert }(-1)^{l}\left( \left\vert
j,2l-1\right\rangle -i\left\vert j,2l\right\rangle \right) ],
\label{solution2}
\end{eqnarray}%
with$\ A_{k_{x}}=\left( v-w\right) /\left( J_{x}\sin k_{x}\right) $ the $%
k_{x}$-dependent amplified amplitude.

Two solutions have the following implications when we consider incidences
with opposite helicities. (i) $\left\vert \psi _{k_{x}}^{L}\right\rangle $
indicates a perfect resonant transmission for the $\left( \left\vert
j,2l-1\right\rangle -i\left\vert j,2l\right\rangle \right) $-type\
(left-handed) incident wave, where $-i$ indicates the momentum $k_{y}=-\pi
/2 $. (ii) $\left\vert \psi _{k_{x}}^{R}\right\rangle $ indicates a
combination of perfect resonant transmission and an equal amplitude $\left(
\left\vert j,2l-1\right\rangle -i\left\vert j,2l\right\rangle \right) $%
-type\ out going waves with $k_{x}$-dependent amplitude for the $\left(
\left\vert j,2l-1\right\rangle +i\left\vert j,2l\right\rangle \right) $-type
(right-handed)\ incident wave, where $+i$ indicates the momentum $k_{y}=+\pi
/2$. The amplitude is inversely proportional to the group velocity $%
2J_{x}\sin k_{x}$\ of the incidence, and diverges only for zero group
velocity.\ In the region $j>0$, we have%
\begin{equation}
\left\vert \langle j,2l-\lambda \left\vert \psi _{k_{x}}^{R}\right\rangle
\right\vert ^{2}=\frac{1}{2M}\left[ \left( -1\right) ^{\lambda }+\frac{w-v}{%
J_{x}\sin k_{x}}\right] ^{2},  \label{pattern}
\end{equation}%
which indicates an interference pattern along the transverse direction ($y$
direction). In particular, for $\sin k_{x}=\pm (w-v)/J_{x}$, it exhibits
transverse standing-wave mode, which indicates that the transmitted waves
only appear in the tube leads embedded the blue sites or the yellow sites.

So far we have given a complete analysis of the scattering problem for the
incidence with transverse wave vector $k_{y}=\pm\pi /2$. For other
incidences with $k_{y}\neq \pm\pi /2$, the coalescing state of the
non-Hermitian SSH ring is a\ forbidden channel due to the mismatch of the
transverse momentum\ $k_{y}$. Other channel in the scattering center can
also be eliminated adiabatically at large staggered coupling $w,\nu \gg
J_{x},J_{y}$, resulting in near perfect reflection. As a temporary summary,
we can conclude that the present system allows the coexistence of two types
of helical resonant transport: (i) zero transverse bandwidth perfect
transmission, (ii) zero transverse bandwidth amplified transmission and
reflection. These phenomena are schematically simulated in Fig. \ref{figure3}%
.

To verify and demonstrate the performance of the proposed scheme, we
simulate the time evolutions for both two typical cases. We consider two
kinds of Gaussian wave packet with opposite helicities as the initial
excitation
\begin{equation}
\left\vert \Psi _{\mathrm{G}}^{\pm }(0)\right\rangle =\Omega ^{-\frac{1}{2}%
}\sum_{j=-\infty }^{\infty }\sum_{l=1}^{2M}e^{-\frac{\alpha _{\mathrm{w}%
}^{2}\left( j-N_{\mathrm{c}}\right) ^{2}}{2}}e^{ik_{x}j}e^{\pm i\frac{\pi }{2%
}l}\left\vert j,l\right\rangle ,  \label{G}
\end{equation}%
where $\Omega =2M\sqrt{\pi }/\alpha _{\mathrm{w}}$ is the normalization
constant. The parameter $\alpha _{\mathrm{w}}$\ determines the width of the
Gaussian wavepacket and $N_{\mathrm{c}}$\ is the initial position of its
center. We numerically calculate the evolved wave function $\left\vert \Psi
_{\mathrm{G}}^{\pm }(t)\right\rangle =e^{-iHt}\left\vert \Psi _{\mathrm{G}%
}^{\pm }(0)\right\rangle $ for the system with $M=2$ [see Fig. \ref{figure4}%
(a)]. The non-Hermitian SSH ring as the filter is located at the position $%
x=0$. We present the numerical results, including the profiles of the
evolved excitations and the probability distribution in $y$ direction for
the evolved excitations in Figs. \ref{figure4}(b)-(e). We also present the
numerical results for the system with $M=4$ in Fig. \ref{figureB1} (see
Appendix B). Both cases verify our predictions.

\section{Purification and Amplification by the Helical Filter}

\label{dynamics}

\begin{figure}[tbp]
\centering
\includegraphics[width=0.352\textwidth]{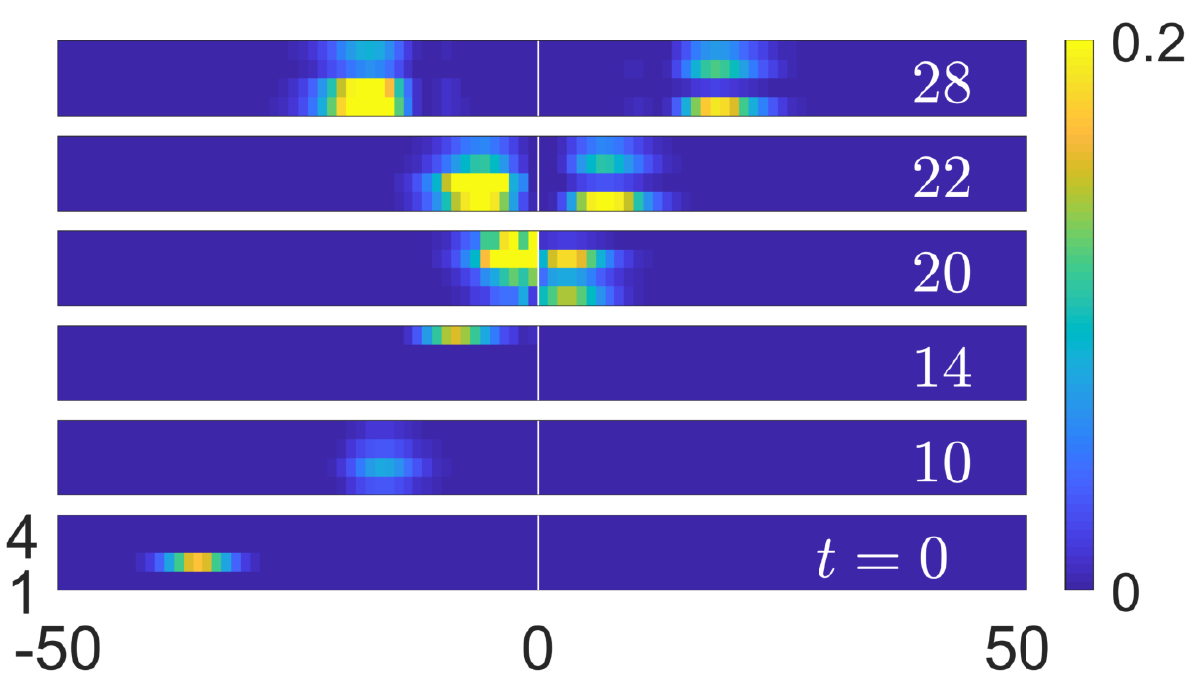}
\caption{Snapshots of the intensity for single-site excitation in the $y$
direction. The parameters of the initial excitation are $\protect\alpha _{\mathrm{w}}=0.3$, $k_{x}=-\protect\pi/2$ and $N_{\mathrm{c}}=-35$, and of
the system are $J_{x}=J_{y}=J=0.25$, $w=1.5$, $v=0.5$. The time unit is $J^{-1}$.}
\label{figure5}
\end{figure}

For the incidence possessing an opposite helicity with the non-Hermitian SSH
ring filter, the filter generates a $(w-v)/\left( J_{x}\sin k_{x}\right) $
times amplified out going wave possessing both identical and opposite
helicities with the filter toward different directions. If $\left(
w-v\right) \gg J_{x}$ or $k_{x}$ approaches to $0$ and $\pi $, the
reflection and transmission beams dominate after the incidence is scattered
at the filter.

We first consider an initial single-site excitation in the $y$ direction in
comparison with a plane wave excitation in Eq. (\ref{G}). The filter has the
left-handed helicity and the initial excitation is Gaussian in the $x$
direction in the form of
\begin{equation}
\left\vert \Psi _{\mathrm{G}}(0)\right\rangle =\Omega ^{-\frac{1}{2}%
}\sum_{j=-\infty }^{\infty }e^{-\frac{\alpha _{\mathrm{w}}^{2}\left( j-N_{%
\mathrm{c}}\right) ^{2}}{2}}e^{ik_{x}j}\left\vert j,2\right\rangle ,
\end{equation}%
where $\Omega =\sqrt{\pi }/\alpha _{\mathrm{w}}$ is the normalization
constant. From the simulation in Fig. \ref{figure5}, we notice that the
evolved excitation is amplified after being scattered at the filter. In the $%
x>0$ region, the light and dark areas as a typical result of interference is
observed. This indicates that the output in this region consists of two
components with opposite helicities. The single-site excitation in the $y$
direction consists all the possible discrete momentum $k_{y}$ determined by
the size $2M$. Consequently, we notice the spreading of single-site
excitation at the moment $t=10$ before reaching the filter. After
scattering, all the components with momentum $k_{y}\neq \pm \pi /2$ are
reflected back to the $x<0$ region because their energies mismatch the
energies of the filter. The component with left-handed helicity ($k_{y}=-\pi
/2$) or right-handed helicity ($k_{y}=\pi /2$) is resonant with the
coalesced zero mode of the filter. The component with left-handed helicity ($%
k_{y}=-\pi /2$) is perfect transmitted. However, the component with
right-handed helicity ($k_{y}=\pi /2$), except for a perfect transmission,
can induces the right-handed reflected and left-handed transmitted wave due
to the gain in the filter; and the right-handed reflected and left-handed
transmitted wave are the dominant components after scattering.

\begin{figure}[tbp]
\centering
\includegraphics[width=0.5\textwidth]{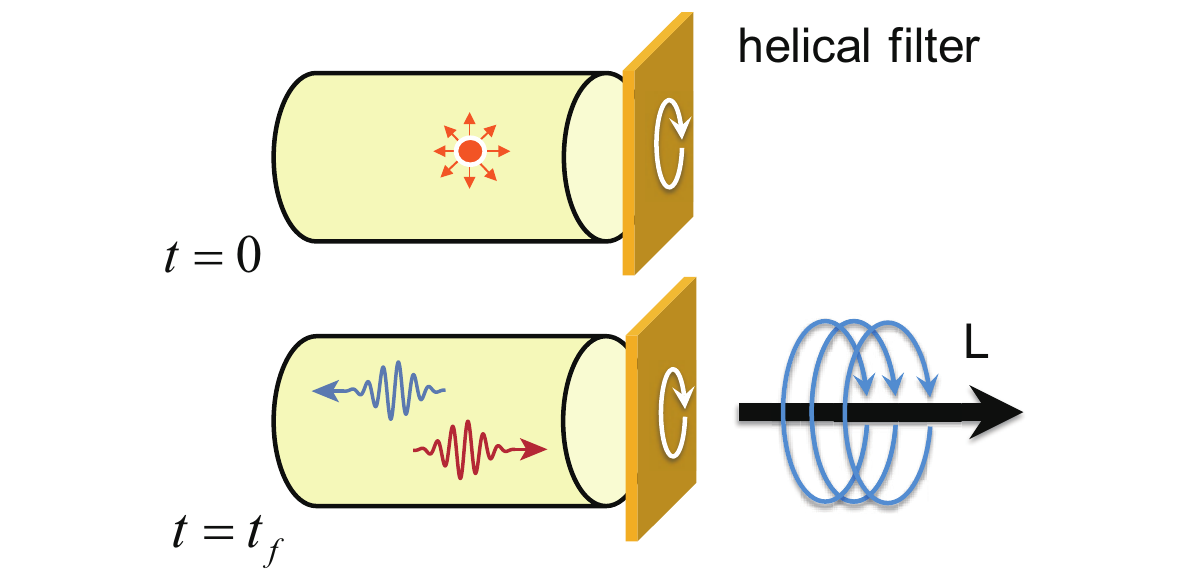}
\caption{Schematic of purified amplification triggered by a single-site
excitation. Both helical filter and passing wave have the left-handed
helicity.} \label{figure6}
\end{figure}

\begin{figure}[tbp]
\centering
\includegraphics[width=0.484\textwidth]{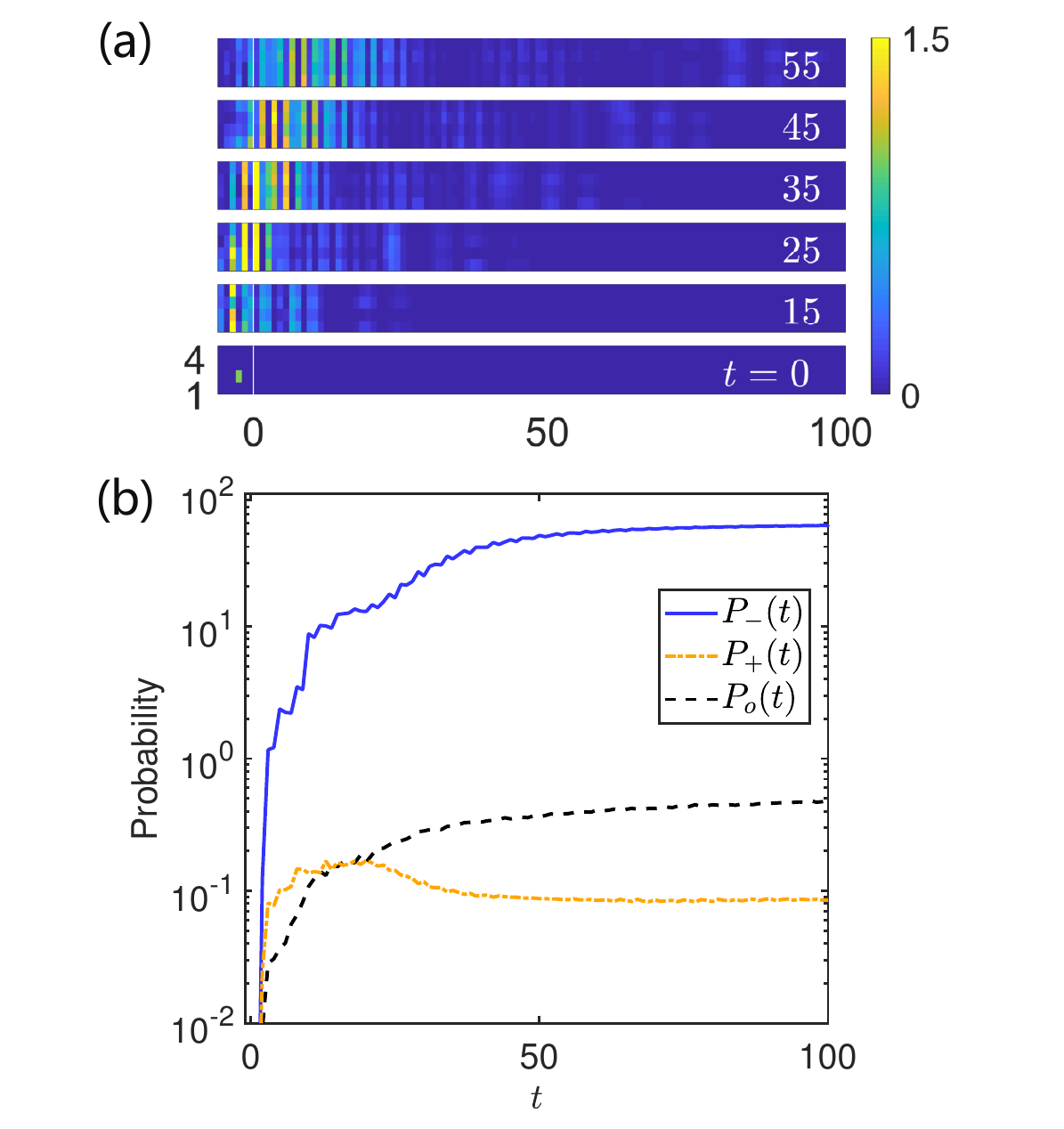}
\caption{(a) Snapshots of the intensity of evolved wave for a single-site
excitation. The dynamics indicates an amplified single helicity output. The
system parameters are $M=2$, $J_{x}=J_{y}=J=0.25$, $w=1.5$, $v=0.5$ and the
time unit is $J^{-1}$. (b) Total transmitted intensity for (a). The blue and
gold lines represent the transmitted intensity ($x>0$) with the left-handed
and right-handed helicities, respectively. The black dash line represents
the summation of intensities for all the other momentum components.} \label%
{figure7}
\end{figure}

Then, we consider a single-site excitation in the tube. The tube has a hard
boundary on its left side (Fig. \ref{figure7}), which perfectly reflects all
the waves. The left boundary together with the filter form a chamber. For a
single-site excitation, the excitation includes the momentum components in
the full region of $k_{x}$ and $k_{y}$. However, only wave with right-handed
or left-handed helicity ($k_{y}=\pm \pi /2$) can pass the filter and are
amplified with one dominant helicity identical to the filter in the output.
The waves with different momentum $k_{x}$ has different amplified ratio and
moving velocity in the $x$ direction. Figure \ref{figure6} schematically
explains the dynamics of a single-site excitation. In Fig. \ref{figure7}, we
consider an initial single-site excitation in the chamber. The profiles of
the evolved wave functions are simulated. The output wave well demonstrates
our description. In Fig. \ref{figure7}(a), the fringes indicate the purified
passing waves with left-handed helicity (being amplified and then being
dominant) similar as the incidence in Figs. \ref{figure4}(b) and \ref%
{figure4}(c); the intensity spreading in the $x$ direction indicates the
output with different velocities. In Fig. \ref{figure7}(b), the intensities
of different helicities as functions of time are depicted. The output after
scattered by the helical filter has a dominant left helicity, which
possesses identical helicity with the filter.

\section{Discussion}

\label{Discussion}

\begin{figure}[tb]
\centering
\includegraphics[width=0.5\textwidth]{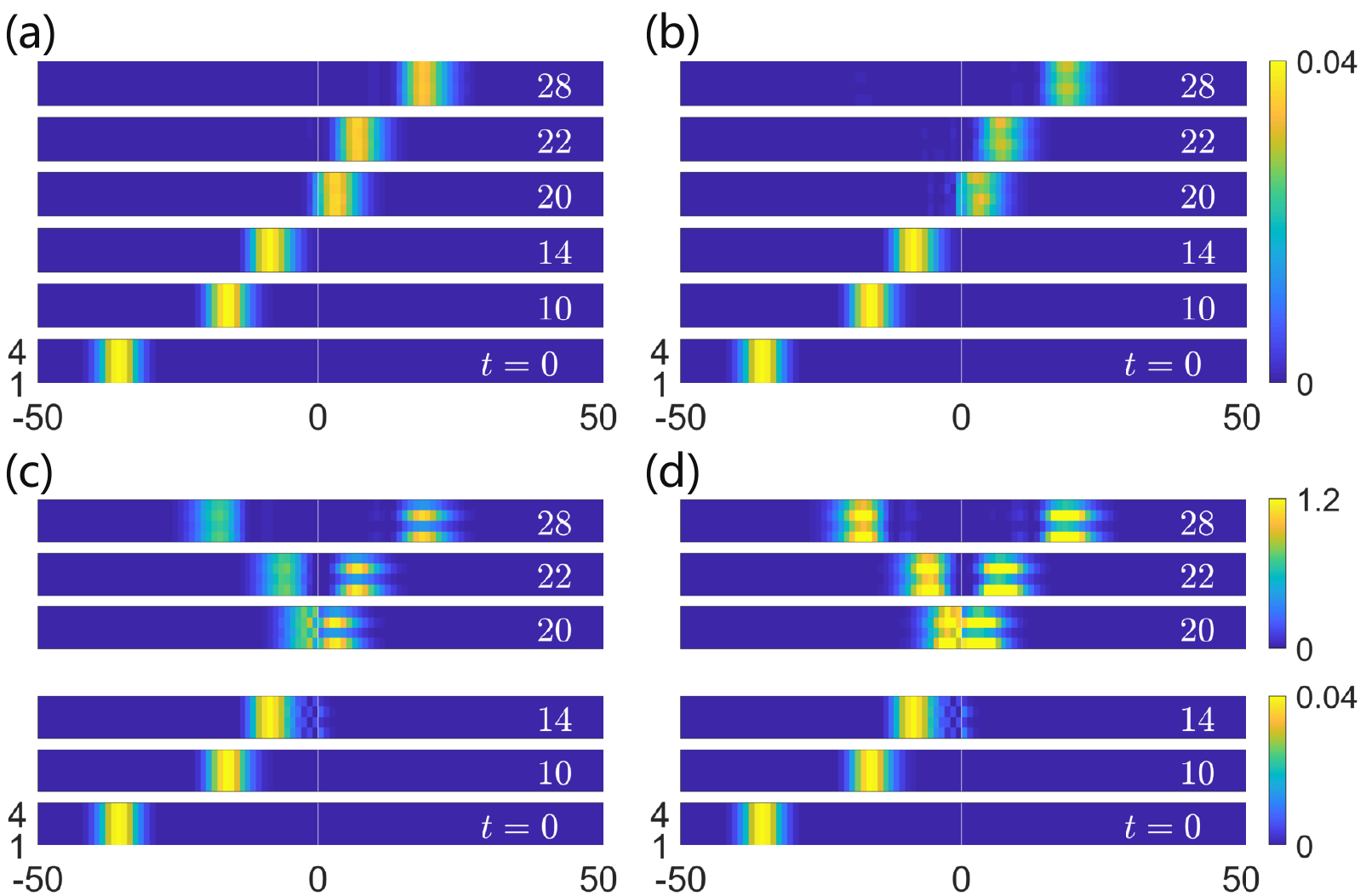}
\caption{Dynamics for $M=2$ system with random disorder in the parameters of the SSH ring. (a, b) Perfect transmission case as comparison with Fig. \ref{figure4}(b).
(c, d) Interfered transmission case as comparison with Fig. \ref{figure4}(c). $R=0.01$ for (a, c) and $R=0.05$ for (b, d). Other parameters are $\protect\alpha _{\mathrm{w}}=0.3$, $k_{x}=-\protect\pi/2$ and $N_{\mathrm{c}
	}=-35$ for the initial excitation and $J_{x}=J_{y}=J=0.25$, $w=1.5$, $v=0.5$ for the system. The unit of time is $J^{-1}$.}
\label{figure8}
\end{figure}

In the previous analytical and numerical calculations, the parameters of SSH
ring in the system is set at the EP exactly. However, in the experiment, the
imperfections of defect or impurity may present; which may drive the
non-Hermitian system unstable with exponential increase of excitation
intensity. Notably, the zero energy of non-Hermitian SSH ring engineered at
the EP is sensitive to the system parameters, and is thus sensitive to the
system imperfections. Thus, it is worth to investigate the dynamics in the
presence of imperfections or at system parameters deviated from the EP. On
the one hand, we consider the influence of imperfections present in all the
parameters of the SSH ring
\begin{eqnarray}
w_{l} &=&w+\delta _{w,l},  \notag \\
v_{l} &=&v+\delta _{v,l}, \\
\gamma _{l} &=&w-v+\delta _{\gamma ,l},  \notag
\end{eqnarray}%
where $\delta _{w,l}$, $\delta _{v,l}$, and $\delta _{\gamma ,l}$ are all
random real numbers within the interval $\left[ -R,R\right] $ and $l$ is the
site index. The numerical simulations for the perfect transmission and
interfered transmission simulated in Fig. \ref{figure4}(b) and (c) under $%
R=0.01$ and $R=0.05$ are shown in Fig. \ref{figure8}. We observe that when $%
R=0.01$, the dynamics of perfect transmission and interfered transmission
are almost unaffected although in the presence of random imperfections;
however, when $R=0.05$, the imperfections apparently affect the wave
dynamics. The numerical results show that the dynamics in the system is not
sensitive to slight imperfections. The reason is that the imaginary part of
the perturbed zero energy is not large enough to exhibit exponential
increase of excitation intensity in a limited time region.

On the other hand, when the parameters of the SSH ring are deviated from the
EP, the lattice support the bound state \cite{Longhiboundstate} with
imaginary energy at the large deviation. We numerically investigate the
system in Fig. \ref{figureB1}(a) (see Appendix B), where the gain and loss
for the system is chosen $\pm i\gamma =\pm i(w-v)=\pm i$. We replace $\gamma
$ by $\gamma \rightarrow \gamma +\delta $ with a positive small real number $%
\delta $. The eigenstates of the zero energy levels for the system with $%
\delta =0$ are all extended states. Some of them are the coalesced zero
energy levels, which are responsible for the helical dynamics. When the
deviation from EP is slight, for example $\delta \leqslant 0.01$, though the
imaginary energy levels appear, the bound state does not exist and the
helical dynamics are almost unaffected. When $\delta =0.15$, the bound
states localized at the SSH ring with imaginary energy are observed. The
absence of bound state with imaginary energy explains why the system is
stable under slight imperfections.

The 2D square lattice tube is not easily to be realized. 
However, the experimental realization of quasi-1D systems shown 
in Fig. \ref{figure1}(b) is ready. The quasi-1D system is essential for the
helical dynamics proposed in the square lattice tube. We discuss a possible
realization with the coupled resonator array. The quasi-1D system considered
in Fig. \ref{figure1}(b) at $k_{y}=\pi /2$ consists of two uniformly coupled
resonator arrays and a $\mathcal{PT}$-symmetric non-Hermitian dimer
scattering center with gain and loss $\pm i\left( w-v\right) $; which can be
induced by pumping the doped ions and sticking additional absorption
material. The single inter-array nonreciprocal coupling (represented by the
red arrow) has a Peierls phase factor $e^{\pm i\pi /2}$ in the front. The
Peierls phase can be generated through a path length imbalance method \cite%
{Hafezi}. From the scattering solutions in Eqs. (\ref{sol_L}) and (\ref%
{sol_R}) (see Appendix A), we can see that the quasi-1D system support the
perfect transmission and interfered transmission; while it can not be used
to observe the purified amplification dynamic proposed in the 2D system. The
non-Hermitian SSH model is a prototypical topological model, further study
on the application of other $\mathcal{PT}$-symmetric non-Hermitian
topological system \cite{LonghiPRL,HJiang,LonghiAAH} and topological system
with non-Hermitian skin effect \cite{CHLee,LonghiSE} will be interesting.
All these 1D and quasi-1D systems are ready to be realized in the
experiments \cite{Helbig,Hofmann}.

The helical dynamics can be observed in a more simple 2D square lattice with
$J_{y}=0$ in Fig. \ref{figure1}(a), which is a series of uniform chains
coupled in the $y$ direction only through the non-Hermitian SSH chain at the
center. Notably, although the scattering center is a non-Hermitian SSH chain
embedded in the simple 2D square lattice, good quality helical dynamics can
be observed in the numerical simulations, which are shown in Fig. \ref{figureB2} in the Appendix B. It is better to
choose large site number in the $y$ direction because that the SSH chain is
more close to the SSH ring under large site number limitation.

\section{Summary}

\label{Summary}

In summary, we propose the helical transport controlled by a $\mathcal{PT}$%
-symmetric non-Hermitian SSH ring at the EP as a helical filter. The filter
has a chiral coalescence state due to the chirality of EP; therefore, the
filter only ensures the resonant transmission for the incidence possessing
an identical helicity with it. For the incidence possessing an opposite
helicity with the filter, except for a full transmission; additional
amplified transmission possessing identical helicity and reflections
possessing opposite helicity with the filter are stimulated. The
transmissions with opposite helicities interfere and create an interference
pattern, which depends on the velocity of the incidence. We first propose a
helical incidence dependent scattering and discuss the multiple-channel
scattering problem in a two-dimensional non-Hermitian square lattice. The
non-Hermitian SSH ring as a helical filter purifies the excitation: the
amplified component in the transmission possesses identical helicity with
the filter and can be dominates, associated with a reflection of opposite
helicity. Our findings provide an application of the non-Hermitian SSH
system and are valuable for the design of optical device using non-Hermitian
metamaterial.

\section*{Appendix}

\setcounter{equation}{0} \renewcommand{\theequation}{A\arabic{equation}} %
\setcounter{figure}{0} \renewcommand{\thefigure}{B\arabic{figure}} %
\renewcommand{\thesubsection}{\Alph{subsection}}

\subsection{Bethe ansatz solution}

In this Appendix, we present the derivations of the scattering solutions in
Eqs. (\ref{solution1}) and (\ref{solution2}), which is the heart of this
work. We first focus on the scattering solutions of the Hamiltonian in Eq. (%
\ref{H_pi2}). The Bethe ansatz wave function of a scattering state $%
\left\vert \psi _{k_{x}}\right\rangle $ has the form
\begin{equation}
\left\vert \psi _{k_{x}}\right\rangle =\sum_{j=-\infty }^{\infty }\left(
f_{j}^{\alpha }\left\vert j,\pi /2\right\rangle _{\alpha }+f_{j}^{\beta
}\left\vert j,\pi /2\right\rangle _{\beta }\right) .
\end{equation}%
The Schr\"{o}dinger equation $H_{\pi /2}\left\vert \psi
_{k_{x}}\right\rangle =E_{k_{x}}\left\vert \psi _{k_{x}}\right\rangle $
gives
\begin{eqnarray}
J_{x}\left( f_{j-1}^{\alpha }+f_{j+1}^{\alpha }\right)
&=&E_{k_{x}}f_{j}^{\alpha },j\neq 0,  \notag \\
J_{x}\left( f_{j-1}^{\beta }+f_{j+1}^{\beta }\right)
&=&E_{k_{x}}f_{j}^{\beta },j\neq 0,  \notag \\
J_{x}f_{1}^{\alpha }+J_{x}f_{-1}^{\alpha }+2\left( w-v\right) f_{0}^{\beta }
&=&E_{k_{x}}f_{0}^{\alpha },  \notag \\
J_{x}\left( f_{1}^{\beta }+f_{-1}^{\beta }\right) &=&E_{k_{x}}f_{0}^{\beta }.
\label{eq_f}
\end{eqnarray}

Considering an incident plane wave with momentum $k_{x}$ incoming from one
side of lead $\alpha $, the ansatz wave function $f_{j}^{\alpha /\beta }$
has the form%
\begin{eqnarray}
f_{j}^{\alpha } &=&\left\{
\begin{array}{cc}
e^{ik_{x}j}+r_{k_{x}}^{\alpha }e^{-ik_{x}j}, & j\leqslant -1 \\
t_{k_{x}}^{\alpha }e^{ik_{x}j}, & j\geqslant 0%
\end{array}%
\right. ,  \notag \\
f_{j}^{\beta } &=&\left\{
\begin{array}{cc}
r_{k_{x}}^{\beta }e^{-ik_{x}j}, & j\leqslant -1 \\
t_{k_{x}}^{\beta }e^{ik_{x}j}, & j\geqslant 0%
\end{array}%
\right. .  \label{f_alpha}
\end{eqnarray}%
Here $r_{k_{x}}^{\alpha /\beta }$ and $t_{k_{x}}^{\alpha /\beta }$ are the
reflection and transmission amplitudes of the incident wave. Substituting
Eq. (\ref{f_alpha}) into Eq. (\ref{eq_f}), we obtain the energy
\begin{equation}
E_{k_{x}}=2J_{x}\cos k_{x},
\end{equation}%
and the reflection and transmission amplitudes
\begin{eqnarray}
t_{k_{x}}^{\alpha } &=&1,r_{k_{x}}^{\alpha }=0,  \notag \\
t_{k_{x}}^{\beta } &=&r_{k_{x}}^{\beta }=0.
\end{eqnarray}%
Then the scattering wave function is
\begin{equation}
\left\vert \psi _{k_{x}}^{L}\right\rangle =\sum_{j=-\infty }^{\infty
}e^{ik_{x}j}\left\vert j,\pi /2\right\rangle _{\alpha }.  \label{psi_1}
\end{equation}

For the case of an incident plane wave with momentum $k_{x}$ incoming from
one side of lead $\beta $, we set $f_{j}^{\alpha /\beta }$ as the form
\begin{eqnarray}
f_{j}^{\alpha } &=&\left\{
\begin{array}{cc}
r_{k_{x}}^{\alpha }e^{-ik_{x}j}, & j\leqslant -1 \\
t_{k_{x}}^{\alpha }e^{ik_{x}j}, & j\geqslant 0%
\end{array}%
\right. ,  \notag \\
f_{j}^{\beta } &=&\left\{
\begin{array}{cc}
e^{ik_{x}j}+r_{k_{x}}^{\beta }e^{-ik_{x}j}, & j\leqslant -1 \\
t_{k_{x}}^{\beta }e^{ik_{x}j}, & j\geqslant 0%
\end{array}%
\right. .
\end{eqnarray}%
Substituting it into Eq. (\ref{eq_f}), we obtain the energy $%
E_{k_{x}}=2J_{x}\cos k_{x}$ and the reflection and transmission amplitudes
\begin{eqnarray}
r_{k_{x}}^{\alpha } &=&t_{k_{x}}^{\alpha }=\frac{(w-v)i}{J_{x}\sin k},
\notag \\
r_{k_{x}}^{\beta } &=&0,t_{k_{x}}^{\beta }=1.
\end{eqnarray}%
\begin{figure}[tbp]
	\centering
	\includegraphics[width=0.5\textwidth]{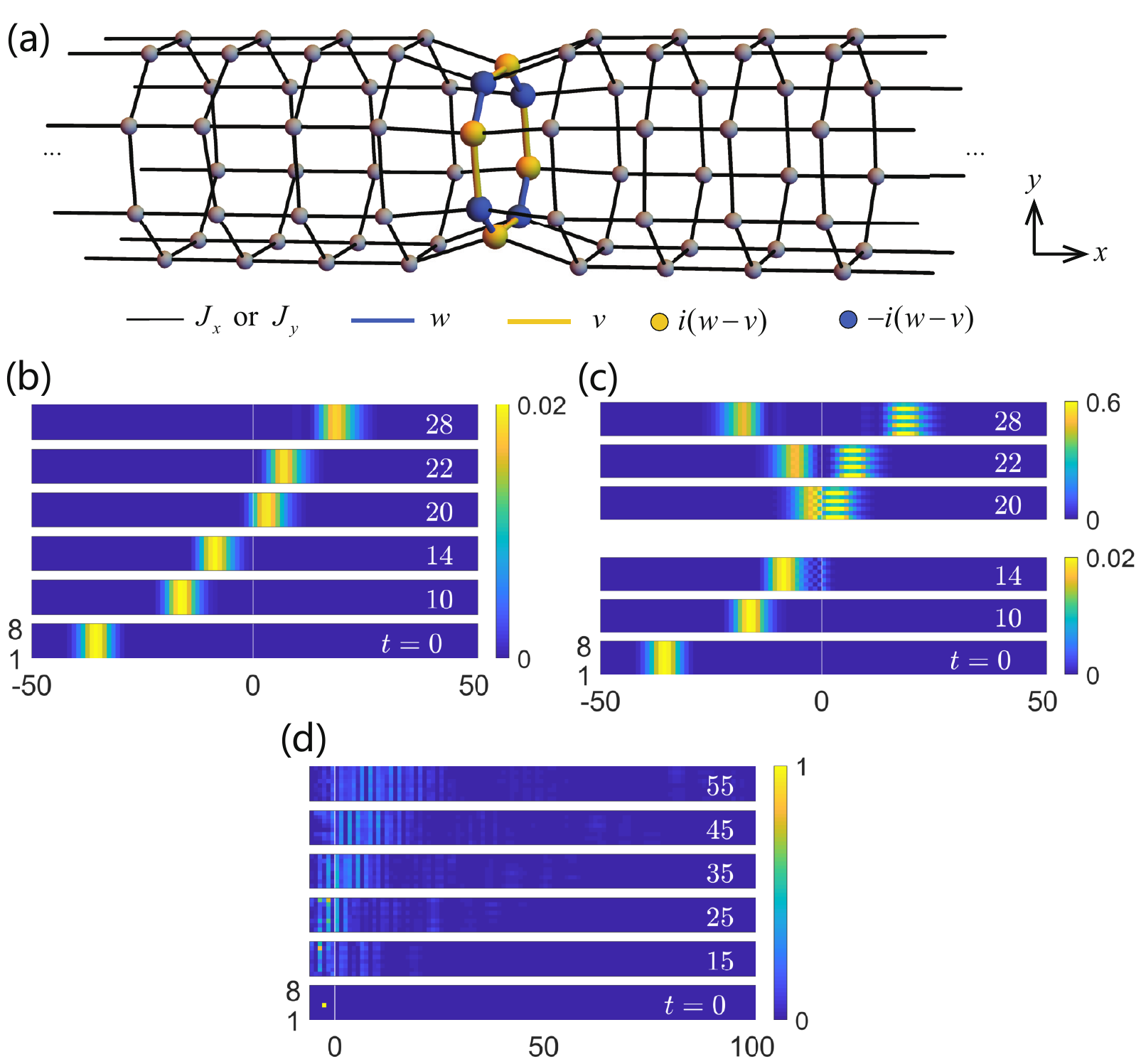}
	\caption{(a) Non-Hermitian lattice tube with $M=4$. Snapshots of the
		intensity at various time moments for three typical initial excitations: (b)
		Perfect transmission of Gaussian profile $\left\vert \Psi _{\mathrm{G}
		}^{-}(0)\right\rangle$ with identical helicity of the filter; (c) interfered
		transmission of Gaussian profile $\left\vert \Psi _{\mathrm{G}		}^{+}(0)\right\rangle$ with opposite helicity of the filter; (d)
		purification and amplification of the single-site excitation. The
		interference pattern in (c) is in accords with the result in Eq. (\protect
		\ref{pattern}), containing four light and four dark spots. Other parameters
		of the initial excitation and system are identical with the $M=2$ case in
		Figs. \protect\ref{figure4} and \protect\ref{figure7}. } \label{figureB1}
\end{figure}
Then the scattering wave function is
\begin{equation}
\left\vert \psi _{k_{x}}^{R}\right\rangle =\sum_{j=-\infty }^{\infty }\left[
\frac{(w-v)i}{J_{x}\sin k_{x}}e^{ik_{x}\left\vert j\right\vert }\left\vert
j,\pi /2\right\rangle _{\alpha }+e^{ik_{x}j}\left\vert j,\pi /2\right\rangle
_{\beta }\right] ,  \label{psi_2}
\end{equation}

After applies the inverse transformation of Eq. (\ref{tran_1}), the above
two scattering solutions Eqs. (\ref{psi_1}) and (\ref{psi_2}) can be written
as
\begin{equation}
\left\vert \psi _{k_{x}}^{L}\right\rangle =\frac{1}{\sqrt{2}}\sum_{j=-\infty
}^{\infty }e^{ik_{x}j}\left( \left\vert j,\pi /2\right\rangle
_{1}-\left\vert j,\pi /2\right\rangle _{0}\right) ,  \label{sol_L}
\end{equation}%
and
\begin{eqnarray}
\left\vert \psi _{k_{x}}^{R}\right\rangle &=&\frac{-i}{\sqrt{2}}%
\sum_{j=-\infty }^{\infty }\left[ e^{ik_{x}j}\left( \left\vert j,\pi
/2\right\rangle _{1}+\left\vert j,\pi /2\right\rangle _{0}\right) \right]
\notag \\
&&-\frac{w-v}{J_{x}\sin k_{x}}e^{ik_{x}\left\vert j\right\vert }\left(
\left\vert j,\pi /2\right\rangle _{1}-\left\vert j,\pi /2\right\rangle
_{0}\right) ],  \label{sol_R}
\end{eqnarray}%
in the space of Hamiltonian Eq. (\ref{H_pi1}). Applying the inverse Fourier
transformations with $k_{y}=\pi /2,$ we obtain the solutions in real space

\begin{equation}
\left\vert \psi _{k_{x}}^{L}\right\rangle =\frac{-i}{\sqrt{2M}}%
\sum_{j=-\infty }^{\infty }\sum_{l=1}^{M}(-1)^{l}e^{ik_{x}j}\left(
\left\vert j,2l-1\right\rangle -i\left\vert j,2l\right\rangle \right) ,
\end{equation}%
and
\begin{eqnarray}
\left\vert \psi _{k_{x}}^{R}\right\rangle &=&\frac{-1}{\sqrt{2M}}%
\sum_{j=-\infty }^{\infty }\sum_{l=1}^{M}(-1)^{l}[e^{ik_{x}j}\left(
\left\vert j,2l-1\right\rangle +i\left\vert j,2l\right\rangle \right)  \notag
\\
&&-\frac{w-v}{J_{x}\sin k_{x}}e^{ik_{x}\left\vert j\right\vert }\left(
\left\vert j,2l-1\right\rangle -i\left\vert j,2l\right\rangle \right) ].
\end{eqnarray}

\begin{figure}[t]
	\centering
	\includegraphics[width=0.5\textwidth]{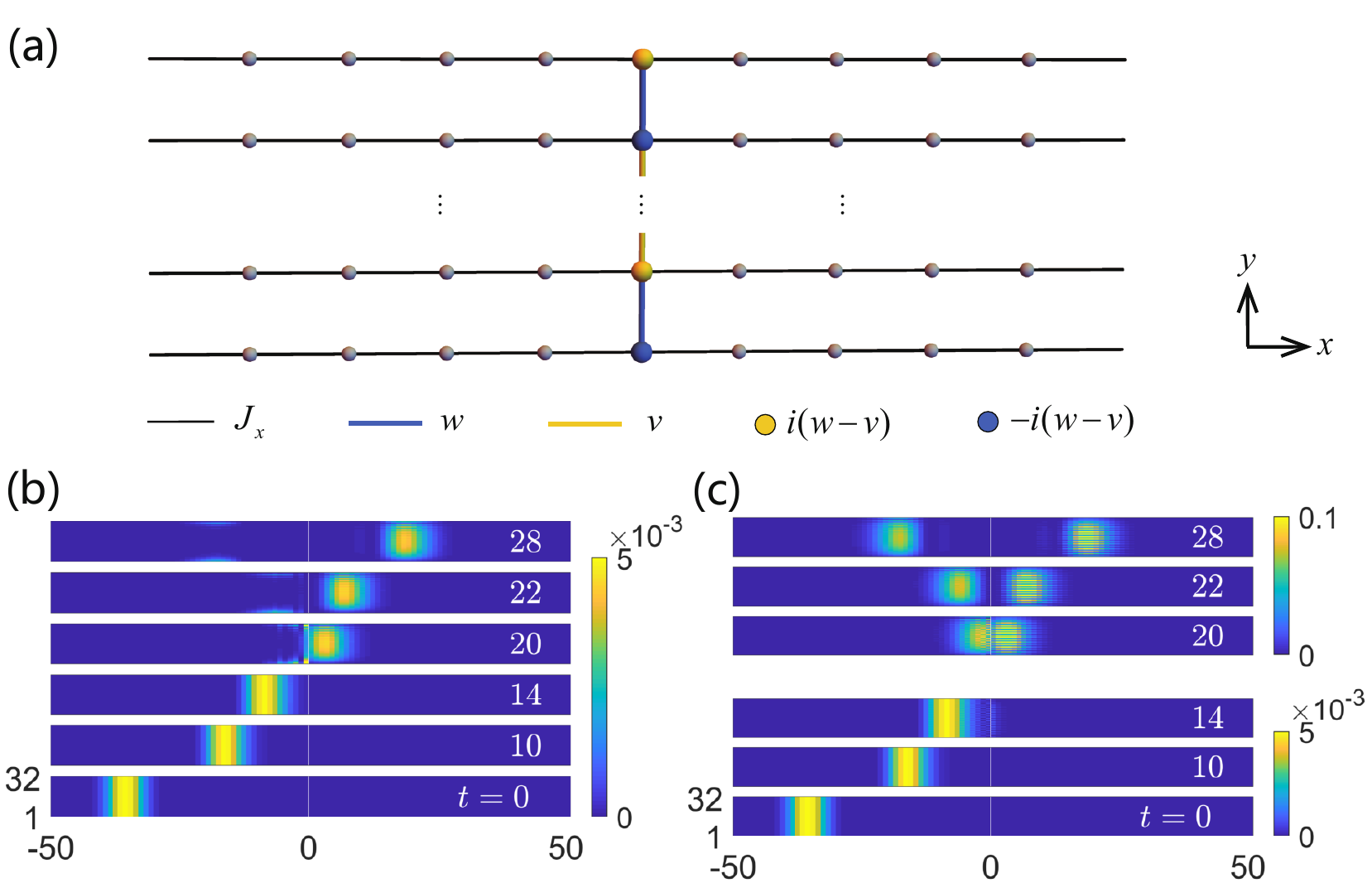}
	\caption{(a) Schematic of the non-Hermitian 2D square lattice. Snapshots of the intensity at various time moments for two typical initial excitations in the 2D square lattice with $M=16$. (b) Perfect transmission of Gaussian profile $\left\vert \Psi _{\mathrm{G} }^{-}(0)\right\rangle$ with identical helicity of the filter; (c) interfered transmission of Gaussian profile 
		$\left\vert \Psi _{\mathrm{G}}^{+}(0)\right\rangle$ with opposite helicity of the filter. Other parameters are $\protect\alpha _{\mathrm{w}}=0.3$, $k_{x}=-\protect\pi/2$ and $N_{\mathrm{c}
		}=-35$ for the initial excitation and $J_{x}=0.25$, $w=1.5$, $v=0.5$ for the system. The unit of time is $J_{x}^{-1}$.} \label{figureB2}
\end{figure}

\subsection{Numerical results for tube with $M=4$ and square lattice with $M=16$}

In Fig. \ref{figureB1}, we present the numerical simulations for the non-Hermitian lattice tube with size $M=4$. The dynamics of perfect transmission, interfered
transmission, and purification and amplification for single-site excitation
are shown.

In Fig. \ref{figureB2}, we present the numerical simulations for the non-Hermitian 2D square lattice with size $M=16$. The dynamics of perfect transmission and interfered transmission are shown.

\acknowledgments We acknowledge the support of National Natural Science
Foundation of China (Grants No. 11874225, No. 11975128, and No. 11605094).

\end{document}